\documentclass[vecphys]{svmult}

\usepackage{makeidx}         % allows index generation
\usepackage{graphicx}        % standard LaTeX graphics tool
\usepackage{multicol}        % used for the two-column index
\usepackage[bottom]{footmisc}% places footnotes at page bottom
\makeindex             % used for the subject index
\begin{document}

\title*{NIBLES: an H{\large I} census of local SDSS galaxies}
\author{W. van Driel\inst{1}, S. Schneider\inst{2}, M. Lehnert\inst{1} 
\and the NIBLES Consortium}
\institute{Observatoire de Paris, GEPI, 92195 Meudon, France
\texttt{wim.vandriel@obspm.fr}
\and Univ. of Massachusetts, Astronomy Program, Amherst, MA 01003, U.S.A. \texttt{}}
\maketitle

{\bf Abstract:} NIBLES (Nan\c{c}ay Interstellar Baryons Legacy Extragalactic Survey) is a Key Project 
proposed for the 100m-class Nan\c{c}ay Radio Telescope (NRT) in France (e.g., Monnier Ragaigne et al. 2003). 
Its aim is a census of the H\,{\sc i} gas content and dynamics of 4,000 Sloan Digital Sky Survey 
galaxies in the Local Volume (900$<$cz$<$12,000 km/s). 
The galaxies were selected based on their total stellar mass 
(absolute z-band magnitude $M_z$), and are distributed evenly over the entire range of 
$M_z$ covered by local SDSS galaxies  (-10 to -24 mag, for H$_0$=70 km\,s$^{-1}$\,Mpc$^{-1}$). 
A related  survey (TICLES) was proposed for CO(1-0) line observations at the IRAM 30m. 
An NRT pilot survey is being made of over 600 galaxies.
NIBLES will be complementary to the ALFALFA and EBHIS blind H\,{\sc i} surveys, which will detect
a different ensemble of local galaxies, and which our pilot survey results indicate will detect 
$\sim$40-45\% of the NIBLES sample.
NIBLES is an open collaboration and anyone interested in the science and willing to 
contribute to the project is welcome to join the score of NIBLErS. \\

\noindent
Science goals of this project are: \\ 

\noindent
Provide a complete and statistically robust analysis of the density of baryons in the local 
Universe, and determine the phase in which these baryons reside.  That is, determine the 
co-moving space density of H$_2$,  H\,{\sc i}, stellar mass, and dynamical mass; 
\noindent

Determine the  H\,{\sc i} Mass Function and the CO luminosity function and the joint probability 
distribution of  H\,{\sc i} mass/CO luminosity and other galaxy characteristics (i.e., infrared 
luminosity, morphological type, stellar mass, stellar age, etc.); 
\noindent

Determine the  H\,{\sc i}/CO gas content and fraction as a function of total stellar mass (fit from the 
optical/near-infrared spectral energy distribution and M/L$_z$), dynamical mass (estimated from  H\,{\sc i} 
and CO line widths), morphological type, and average stellar density; 
\noindent

Determine the systematic relationships and variability between dynamical, stellar, and gas masses (e.g., do 
the estimated dynamical masses change in proportion to total baryonic mass?); 
\noindent

Compare galactic gas fractions and metallicity, to determine the effective yields of the largest sample 
of galaxies available and to allow for the most detailed comparison with theoretical models of the evolution 
of galaxies.  This will enable us to determine the relative influence of outflows and infall on the evolution 
of galaxies. We will use metallicities determined from the SDSS data and compare them with the overall 
gas fractions to determine the effective yields as a function of galaxy type, dynamical and stellar masses, 
and stellar mass surface densities. Comparing these results with chemical evolution models will provide a 
robust test of our understanding of gas recycling yet performed. \\ 

\noindent
The H\,{\sc i} data that are presently available in the literature are largely insufficient for our 
proposed studies; of the 4000 galaxies in the NIBLES sample, only 13\% have published H\,{\sc i} data, 
half of which concern only the gas-rich spiral galaxies in the highest luminosity range (-20.5$<$$M_z$$<$-23.5).
Also the data that will ultimately be provided by the Arecibo ALFALFA and Effelsberg EBHIS blind surveys 
(Giovanelli et al. 2005; Winkel et al. 2007). will not be sufficient; our ongoing NRT pilot survey indicates 
that they will detect 40-45\% of the NIBLES sample.

NIBLES selection criteria are: (1) high quality SDSS magnitudes, and high quality SDSS optical spectra, (2) 
within the Local Volume (recession velocity of 900$<$cz$<$12,000 km/s), (3) sampling of each 0.5 magnitude 
wide bin in $M_z$, with a maximum of 200 galaxies for the most populated bins, and of all galaxies (minimum 20)
for the poorer bins, (4) focus on the most nearby objects in each $M_z$ bin, as these will have the highest  
H\,{\sc i} flux densities. \\ 

\noindent
Pilot surveys: in 2007 we obtained 500 hours of telecope time at Nan\c{c}ay for NIBLES pilot surveys. 
To date, 340 galaxies have been observed for 30 minutes (half the intended average 
time per object in NIBLES). The data have a typical rms noise level of 3.5 mJy at a 10 km/s velocity resolution,
comparable to the point source detection sensitivity of the blind ALFALFA and EBHIS surveys.
The overall detection rate is 66\%, and similar over the entire 12 magnitude range in $M_z$.
We estimate that $\sim$40-45\% of the NIBLES sample galaxies would be detectable by ALFALFA or EBHIS,
based on the sensitivities of these surveys and our detected H\,{\sc i} fluxes.

This shows that NIBLES, which will on average have a two times longer integration time than the NRT pilot 
surveys, will provide a unique H\,{\sc i} census of optically selected Local Volume galaxies.
NIBLES will be complementary to the ALFALFA and EBHIS blind H\,{\sc i} surveys of galaxies in the Local Volume. 
In general, blind H\,{\sc i} surveys provide a basis for determining whether there are classes of galaxies 
being missed in an optically selected sample, such as gas-rich Low Surface Brightness objects, but the 
NRT data will be better for detailed studies of the properties of the SDSS galaxies themselves.

%%%%%%%%%%%%%%%%%%%%%%% referenc.tex %%%%%%%%%%%%%%%%%%%%%%%%%%%%%%

\end{document}